\documentclass[aps,prb,twocolumn,superscriptaddress,showpacs,floatfix]{revtex4-1}
\usepackage{graphicx}
\usepackage{rotating}
\usepackage{mathtools}
\usepackage[normalem]{ulem}

\begin{document}
\title{Free-energy landscapes in magnetic systems from metadynamics}
\date{\today}
\author{Jaroslav T\'{o}bik}
\email{jaroslav.tobik@savba.sk}
\affiliation{Institute of Electrical Engineering, Slovak Academy of Sciences,
D\'{u}bravsk\'{a} cesta 9, SK-841 04 Bratislava, Slovakia}
\author{Roman Marto\v{n}\'{a}k}
\affiliation{Department of Experimental Physics, Faculty of Mathematics, Physics and Informatics, Comenius University in Bratislava, Mlynsk\'{a} dolina F2, 84248 Bratislava, Slovakia}
\author{Vladim\'{\i}r Cambel}
\affiliation{Institute of Electrical Engineering, Slovak Academy of Sciences,
D\'{u}bravsk\'{a} cesta 9, SK-841 04 Bratislava, Slovakia}

\begin{abstract}
Knowledge of free energy barriers separating different states is
critically important for assessment of long-term stability of
information stored in magnetic devices. This information, however, is
not directly accessible by standard simulations of microscopic models
because of the ubiquitous time-scale problem, related to the fact that
the transitions among different free energy minima have character of
rare events.  Here we show that by employing the metadynamics
algorithm based on suitably chosen collective variables, namely
helicity and circulation, it is possible to reliably recover the free
energy landscape. We demonstrate the effectiveness of the new approach on the
example of vortex nucleation process in magnetic nanodot with lowered
spatial symmetry. With the help of reconstructed free energy surfaces (FES)
we show the origin of the symmetry broken vortex nucleation, where one
polarity of the nucleated vortex core is preferred,
even though only in-plane magnetic field is present.
\end{abstract}

\maketitle
Practical application of magnetic devices to information storage
requires long-time stability of the magnetized state, robust with
respect to thermal fluctuations. This can only be guaranteed if the
barriers separating the different free-energy minima are much larger
than the thermal energy. For design of new magnetic storage devices it
is thus critically important to calculate the free energy landscape and
determine the barriers. This information, however, is not easy to
extract from the microscopic models, because of the well-known time
scale problem. While the microscopic dynamics of the magnetization can
be directly followed by time integration of the equations of motion,
the transitions among different free energy minima have character of
rare events and therefore their direct simulation is not
computationally feasible. Very similar problem is often encountered in
other branches of physics and chemistry, where e.g. chemical
reactions, phase transitions and protein folding also represent
thermally activated processes in which the transitions between different free
energy minima are only possible via crossing of a barrier. Traditional 
approach to the problem, commonly applied in the field of chemical reactions, 
is provided by the transition state theory \cite{Pelzer, Kramers, Langer}.
It was shown that this approach is also applicable to magnetic systems 
\cite{Neel, Brown, Braun}, for a recent review see Ref.\cite{Coffey}.
For application of this method it is necessary to determine the height of the 
energy barrier defined as energy difference between the initial state and the highest 
point on the minimum energy path (MEP) connecting the initial and final configurations. 
A practical method to find the MEP and localize the highest saddle point is the 
Nugded Elastic Band (NEB) method \cite{NEB, NEB1}. In order to apply these techniques 
to magnetic systems it is convenient to work in curvilinear coordinates 
and the corresponding version of the original method was proposed in 
Ref.\cite{GNEB-CompPhysComm} as Geodesic Nugded Elastic Band (GNEB) method. 

Another class of methods does not aim at finding the transition path directly 
in the full configuration space but applies first a dimensionality reduction 
by introducing a suitable set of collective variables (CV). 
This class includes number of methods, such as e.g. umbrella
sampling\cite{JCompPhys-Torrie}, 
weighted histogram techniques \cite{PRL-Ferrenberg} and metadynamics 
(for review of metadynamics see Ref.\cite{reviews}). 
Here we show that by applying the metadynamics
algorithm\cite{laio-parrinello} with properly chosen CV it
is possible to reliably recover the free-energy landscape at finite
temperature\cite{PRL-BussiLaio}.

The metadynamics algorithm\cite{laio-parrinello} is based on the
identification of a small number of collective variables which are
related to slow degrees of freedom of the system and clearly distinguish the 
initial and final states\cite{Gervasio}. During sampling of
the system which can be performed e.g. by the integration of the equations 
of motion or by Monte-Carlo (MC) sampling, a history-dependent 
biasing term is added to the total energy. Role of this time dependent 
potential is to discourage the system from re-visiting  
the already visited regions in the space of collective variables. The effect 
of this algorithm is twofold. First,
the barrier surrounding the initial free-energy well is gradually
eliminated until the system is able to escape from it and enter the
basin of attraction of a new state. In this sense, metadynamics provides an acceleration of exploration of the configurational space of the system and allows finding the states which otherwise would be reached only after very long dynamics. Second, after sufficiently long
run filling several minima one is able to recover the underlying
free-energy profile.

The latter aspect, namely the possibility to obtain free energy
landscape as function of few collective variables is very appealing for
our purpose. In particular, the shape of the free energy profile can 
induce
unexpected symmetry breaking mechanism as shown recently on one spin
toy model \cite{SciRep-Tobik}. Another advantage is that the
temperature is treated consistently. 
Different magnetic states could have in principle
different sensitivity to temperature, 
which is correctly taken into account in our method,
thus opening the opportunity to
study lowest energy paths for fixed temperature.
Such situation is more realistic than to study total energy surface in configuration
space at zero temperature. We note that temperature is important parameter 
in exotic structures like skyrmions\cite{PRL-Bogdanov} 
or magnetic bobbers\cite{PRL-Rybakov} and the question of thermal stability of skyrmions
is of considerable current interest.

The level of theoretical framework presented here is micro-magnetism, i.e. classical 
description of magnetization by continuous field $\mathbf{M}$. 
The energy and dynamics of this field is given by
phenomenological Landau-Lifshitz equation:
\begin{eqnarray}
&E&= A\int (\nabla \mathbf{M})^2 d\Omega-\int \mathbf{H_{ext}}\cdot\mathbf{M} d\Omega
\label{eq-Ecomp} \\
&+&\frac{\mu_0}{4\pi} \int \int  \left( \frac{\mathbf{M}\cdot\mathbf{M}'}{|\mathbf{r}-\mathbf{r}'|^3}
-\frac{3\mathbf{M}
\cdot(\mathbf{r}-\mathbf{r}')
\mathbf{M}'\cdot(\mathbf{r}-\mathbf{r}')}{|\mathbf{r}-\mathbf{r}'|^5}
\right) d\Omega d\Omega' \nonumber \\
&\partial_t \mathbf{M}&= -\gamma \mathbf{M}  \times \mathbf{H_{eff}}  - 
    \alpha \frac{\gamma}{M_s} \mathbf{M}\times\left( \mathbf{M} \times \mathbf{H_{eff}}\right) \label{eq-LLG} \\
&\mathbf{H_{eff}}&= -\mathbf{\nabla_M} E. \label{eq-Heff}
\end{eqnarray}  
Here $A$ is exchange constant, $\mathbf{H_{ext}}$ external magnetic field,
$M_s$ is saturation magnetization, $\gamma$ gyro-magnetic moment,
$\alpha$ damping coefficient and $E$ total energy. 
The energy
functional $E$ is composed of exchange energy, Zeeman energy
and stray field energy.
The numerical solution of this problem is based on discretization 
in spatial as well as in time domain.  The limit of acceptable 
discretization is defined by magneto-static exchange length
$l_{\mathrm{exch}}=\sqrt{\frac{2A}{\mu_0M_s^2}}$. On this scale exchange
interaction should prevail over demagnetization energy and therefore
on this scale the magnetization can be considered to be a
constant. Equations (\ref{eq-LLG}) and (\ref{eq-Heff}) describe
magnetization dynamics and will be used later in order to discuss
symmetry broken dynamics.

\begin{figure}[t!]
\includegraphics[width=0.99\columnwidth,keepaspectratio]{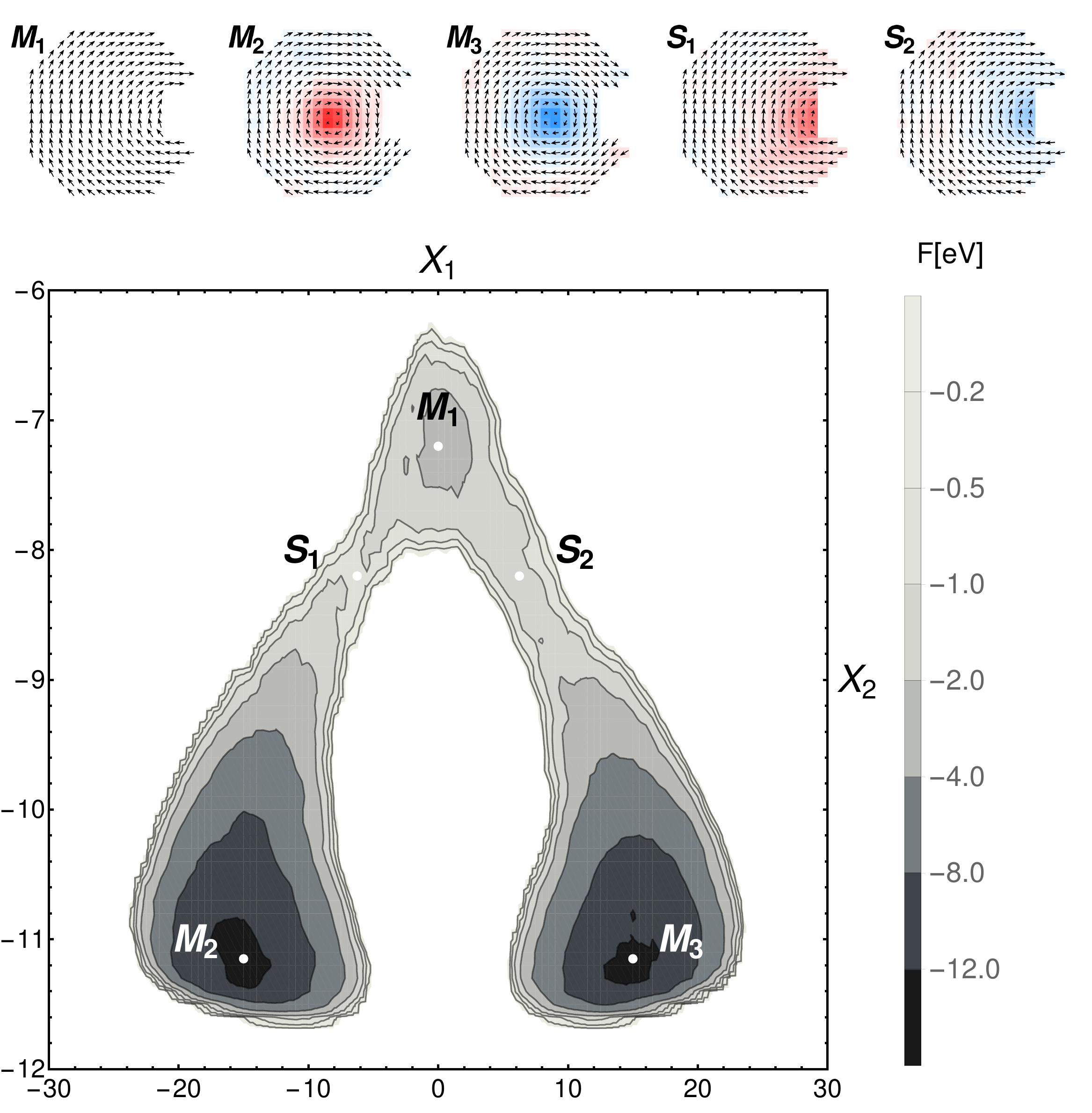}
\caption{Free energy profile as function of collective variables given by 
equations (\ref{eq-colvar1}) and (\ref{eq-colvar2}). The external field
is applied at an angle of $60^\circ$ with respect to axis of symmetry. 
Magnetic configurations
of the initial $C$ state and representative configurations for special 
points are shown in the top panel.
\label{fig-CstateMap}}
\end{figure}

\begin{figure*}[t]
\includegraphics[width=0.99\textwidth,keepaspectratio]{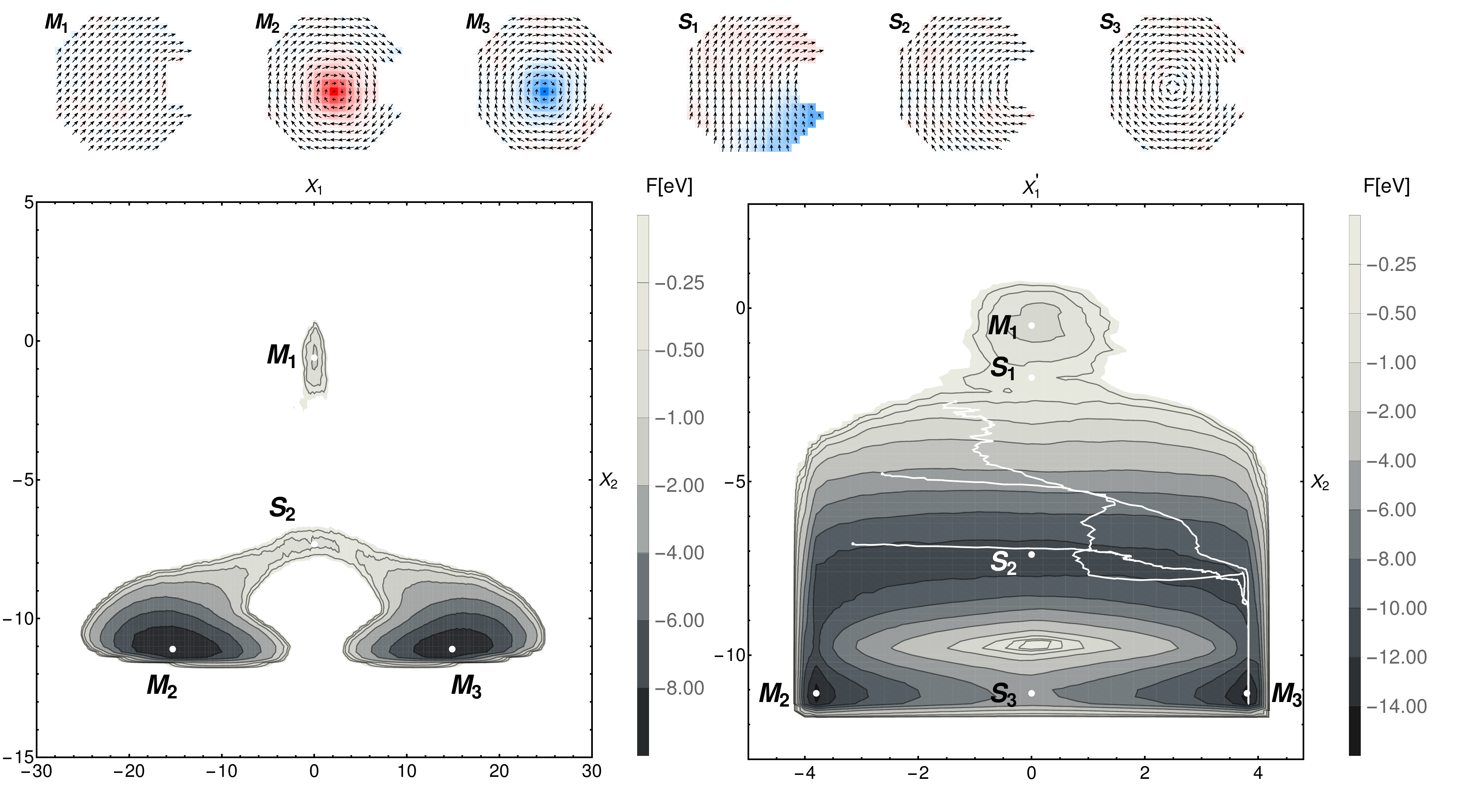}
 \caption{The map of the free energy as function of collective
   variables $X_1$, $X_2$ given by equations
   (\ref{eq-colvar1},\ref{eq-colvar2}) (left panel) and free energy
   as function of the transformed variable  $X_1'$ 
(\ref{eq-vorticityTransformed}) 
   and $X_2$ (right panel). The starting configuration is in the point 
$\mathrm{M_1}$. Vortex 
   state corresponds to global minima $\mathrm{M_2}$ and $\mathrm{M_3}$.
   The lowest saddle between minima
   $\mathrm{M_2}$ and $\mathrm{M_3}$ is labeled $\mathrm{S_2}$. The lowest 
energy barrier
   to climb to escape from local minimum $\mathrm{M_1}$ is defined by saddle point $\mathrm{S_1}$.
   The saddle point $\mathrm{S_3}$ is a numerical artefact related
   to finite discretization of the magnetization field $\mathbf{M}$.
   White color on background indicates un-explored configuration space. 
   Note that region in vicinity of $\mathrm{M_1}$ is not connected to the
   rest of the map on the left panel. 
   \label{fig-SstateMap} 
   } 
\end{figure*}

The choice of the collective variable is crucially important for the
success of the technique. As collective variables we took:
\begin{subequations}
\begin{align}
 X_1 &= \int  M_{z}\left( \partial_x M_{y} - \partial_y M_{x} \right) d\Omega \label{eq-colvar1} \\
 X_2 &= \int  r_{x}(M_{y}-\bar{M}_y) - r_{y}(M_{x}-\bar{M}_x) d\Omega \label{eq-colvar2}
\end{align}
\end{subequations}
Here $\bar{M}_x$ stands for average value $\bar{M}_x=1/\Omega \int M_x d\Omega$. 
The form of collective variables $X_1$
resembles definition of helicity in fluid dynamics. The helicity of
the vector field $\mathbf{v}$ is defined as $\mathbf{v} \cdot \nabla
\times \mathbf{v}$.  The collective variable $X_1$ catches
polarization of the vortex core multiplied by sense of rotation in the
core region. Collective variable $X_2$ reflects circulation of the
spin field and its absolute value is higher for closed circular spin
textures. The negative sign of $X_2$ corresponds to clock-wise rotation
of magnetization in nanodot.  Subtraction of averages in formula
(\ref{eq-colvar2}) guarantees invariance of the $X_2$ variable with respect to
the  
choice of origin for $\mathbf{r}$ coordinate system. Clock-wise
rotation is energetically favored by external field as explained in
our previous work \cite{PRB-Cambel,PRB-Tobik}.  Initial states are characterized by
$X_1=0$, because the initial spin configuration is in-plane for
each spin. The absolute value $|X_2|$ is small for states without vortex. 
The vortex state is characterized by negative value of $X_2$
(because of clockwise rotation of spin field lines 
in vortex states) and positive or negative value 
of $X_1$ depending on polarity of the nucleated vortex. 
It can be seen that the variables $X_1, X_2$ can well distinguish 
between the initial and final states (Figs.~\ref{fig-CstateMap},
\ref{fig-SstateMap}).
Importantly, both CV correspond to slow degrees of freedom since their change requires a massive reorganization of the spin configuration.

Our implementation of metadynamics is based on the Metropolis-Hastings
algorithm of MC sampling method\cite{Metropolis,Hastings}.
The history (time) dependent potential is chosen to have a form of
sum of the Gaussians centered around previously stored collective variables
configurations:
\begin{equation}
V(\mathbf{X})=\sum_{i=1}^T \mathrm{V_0}\, \exp \left(
-\frac{|\mathbf{X}-\mathbf{X_i}|^2}{\sigma_X^2}\right). \label{eq-tdepPot}
\end{equation}
The potential (\ref{eq-tdepPot}) is calculated in each step and is added to
the total energy $E$ in equation (\ref{eq-Ecomp}). The summation index $i$ runs over all stored configurations 
and $\mathbf{X} = (X_1,X_2)$ stands for collective variables. The amplitude of added
Gaussians $V_0$ was set 0.06eV and standard deviation $\sigma_X$ in $(X_1, X_2)$ space
was 0.03 (in arbitrary units after suitable rescaling of the variables, see Figs.~\ref{fig-CstateMap},\ref{fig-SstateMap}).
The time-dependent 
energy enters decision strategy for acceptance of new configuration as in 
standard Metropolis-Hastings algorithm:
\begin{equation}
P(\mathbf{X}_{n+1},\mathbf{X}_n)=\left\{
\begin{array}{ll}
\exp\left(-\frac{E_{n+1}-E_n}{k_B T}\right) \quad &\text{for } E_{n+1}>E_n \\
1 & \text{otherwise}
\end{array}
\right. .
\end{equation}
The local MC steps were realized by changing one spin per step. The
change had to be small in order to keep the acceptance ratio
reasonable. This was done by adding to randomly chosen spin a vector
with amplitude 0.1 and with uniform probability distribution on the
sphere using the Marsaglia algorithm \cite{Marsaglia}. Afterwards the
spin amplitude was renormalized to one. The temperature of T=300K was used
in our simulations.

The values of collective variables were stored with frequency of 100
steps. This choice improves the performance of metadynamics
in several respects. 
It improves diffusivity in the configuration space of collective
variables $\mathbf{X}$,  at the same time allows to use Gaussians
with larger spread value $\sigma_X$, and finally makes the summation 
in equation 
(\ref{eq-tdepPot}) faster \cite{reviews}.  In order to further speed-up 
this summation, we used k-d tree data structure
\cite{kd-tree} to store visited configurations. The k-d tree allows
efficient storage of vectors and subsequent efficient finding of
vectors within prescribed range. The summation of Gaussians in equation
(\ref{eq-tdepPot}) was reduced within rectangle with Manhattan
distance $5\sigma_X$ from the actual value of collective variables
$\mathbf{X}$.

The results of simulation of the modelled system in various magnetic configurations is shown 
in the Figs. \ref{fig-CstateMap} and \ref{fig-SstateMap}.
The nanodot is disk with 70 nm diameter. The removed sector has 1/3 of 
radius with opening angle of
$45^\circ$. The constituting material was Permalloy with saturation magnetization
$M_s=8.6\times 10^5\, A/m$, and exchange constant $A=1.3\times 10^{-11}\,J/m^3$.
The magnetic nanodot was discretized to $N=324$ magnetic moments
$\vec{s}$, each of them having 2 degrees of freedom. This Pacman-like (PL)
nanodot 
was also subject of our previous studies \cite{PRB-Cambel, PRB-Tobik, SciRep-Tobik}.
It was identified earlier that this system has two modes of vortex nucleation\cite{PRB-Tobik}.
One mode is characterized by transition from the C-state (configuration M1 in Fig.
\ref{fig-CstateMap}) to vortex state (configuration M2 or M3). Second mode
is a transition from initial S-state (configuration M1 in Fig. \ref{fig-SstateMap}) to
vortex state (configuration M2 or M3). These two transitions (C-state to vortex or S-state 
to vortex, respectively) differ in the robustness of the vortex polarity control  with respect to out-of-plane
magnetic field\cite{PRB-Tobik} and to temperature\cite{SciRep-Tobik}.

The initial states were prepared by damped zero temperature LLG
dynamics by slowly decreasing external field from high value of
200mT to 30mT (C state) or to 40mT respectively (in case of S state). 
The angle between PL-nanodot axis of symmetry and applied magnetic 
field $\mathbf{H_{eff}}$ is $60^\circ$ for the C-state and $30^\circ$ for the 
S-state.
The value of external magnetic field was fixed during metadynamics simulation.
Initial states had no out-of plane magnetization. The system should undergo 
spontaneous symmetry  breaking (nucleation of vortex) if the external field 
is decreased by 5mT\cite{PRB-Tobik}. The initial states thus represent 
relatively shallow local minima.  

The calculated free energy as function of collective
variables (\ref{eq-colvar1}) and (\ref{eq-colvar2}) for the C-state
is shown in the Fig. \ref{fig-CstateMap}. 
The highest energy point to pass from $\mathrm{M_1}$ 
to minimum $\mathrm{M_2}$ is $\mathrm{S_1}$.
The symmetry related lowest energy path to minimum $\mathrm{M_3}$  
has the highest energy point at the saddle point $\mathrm{S_2}$. 
The symmetry in $X_1$ variable is already broken at the saddle points. 
The height of barrier to overcome from $\mathrm{M_1}$ to $\mathrm{M_2}$ 
or $\mathrm{M_3}$ is 1.8eV, barrier from $\mathrm{M_2}$ or $\mathrm{M_3}$ to
$\mathrm{M_1}$ is 11.8eV. We note that the FES reconstructed with 
the simple version of metadynamics contains some numerical noise 
and the accuracy could be further improved by employing more sophisticated sampling 
techniques such as e.g. well-tempered metadynamics algorithm \cite{PRL_Barducci}.
The time evolution of energy and CV is shown in Fig.~1 in supplemental 
material\cite{supplement}.

In the Fig. \ref{fig-SstateMap} we show the free energy profile
for the metadynamics starting from the S-state ($\mathrm{M_1}$) 
which represents a shallow local minimum. 
The minimum $\mathrm{M_2}$ ($\mathrm{M_3}$)
corresponds to clockwise vortex with positive (negative) polarity 
and the saddle $\mathrm{S_2}$ is C-like state. 
In this figure there are two panels with free energy map.
The left panel shows the local minimum $\mathrm{M_1}$,
disconnected from the rest of the map, since after leaving the high-energy
initial state $\mathrm{M_1}$ the dynamics does not return to this region anymore.
The time evolution
of energy and CV is show in Fig.~2 in supplemental material\cite{supplement}.
The free energy landscape thus can be reliably
reconstructed only in the vicinity of the deep global minima 
$\mathrm{M_2}$, $\mathrm{M_3}$ and the lowest saddle point between
them $\mathrm{S_2}$. Because the full
configuration space including all three minima 
$\mathrm{M_1}, \mathrm{M_2}, \mathrm{M_3}$  
has too big volume we were not able to fill the configuration space 
with Gaussians in reasonable time and connect the minima $\mathrm{M_1}$ 
to $\mathrm{M_2}$ and $\mathrm{M_3}$. Therefore the region of the
starting local minimum $\mathrm{M_1}$ is not on the same energy level as
global minima $\mathrm{M_2}$, $\mathrm{M_3}$.

To circumvent this obstacle we 
decided to use new collective variable $X_1'$ given by equation:
\begin{equation}
 X_1' = 4 \tanh \frac{X_1}{4}. \label{eq-vorticityTransformed} \tag{\ref{eq-colvar1}'}
\end{equation} 
The reason for this transformation is to shrink configuration space
for large values of $X_1$ and thus speed up filling of the global
minima basins. The reconstructed free energy landscape is shown in the 
right panel of Fig.~\ref{fig-SstateMap}. 
The time evolution of energy and transformed CV is shown in Fig.~3 in supplemental
material\cite{supplement}. The topology of the free
energy surface is different with respect to the previous case. 
There are two more saddle points in the free energy map $F(X_1',X_2)$ 
denoted as $\mathrm{S_1}$ and $\mathrm{S_3}$.  The saddle $\mathrm{S_3}$
separates minima $\mathrm{M_2}$ and $\mathrm{M_3}$ in the region of
configuration space with large value of chirality $X_2$. It corresponds to process, 
where the vortex core is shrunk to
zero and then core is created again, but with opposite sign of
helicity. This process is, however, an artifact of the
numerical treatment of the continuous micro-magnetic model and should
not happen in the continuum limit, because the vortex is topologically protected state.
 However, atomic nature of materials permits vortex
destruction by shrinking the vortex core to zero, but the true energy
barrier is probably much higher then our simulation shows.  Note that
a systematic study of the influence of mesh discretization was done in
the case of skyrmion \cite{PRB_Klaui}. The free energy profile along
line $X_1=0$ is shown in the Fig.~4 in Supplemental Material \cite{supplement}. 
The free energy barrier to escape from minimum $\mathrm{M_1}$ via the saddle 
point $\mathrm{S_1}$ is 1.0eV.
Minima $\mathrm{M_2}$ and $\mathrm{M_3}$ are separated by energy barrier 
9.1eV at saddle point $\mathrm{S_2}$. The free energy difference 
between saddle points $\mathrm{S_1}$ and $\mathrm{S_2}$ is 10.2eV.

A peculiar feature of the free energy (Fig. \ref{fig-SstateMap}) map 
is the existence of a common saddle point $\mathrm{S_1}$ which defines 
the lowest energy barrier
between initial in-plane magnetization and both final vortex states.  
The region between saddles $\mathrm{S_1}$ and $\mathrm{S_2}$ is 
region with substantial
energy decrease without necessity to break the symmetry in $X'_1$
variable. When the system reaches saddle point $\mathrm{S_2}$, further
decrease of the energy is possible only if the symmetry of the
collective variable $X'_1$ is broken and the vortex polarity chooses
sign. It was argued earlier that this kind of topology of the energy landscape induces 
breaking of symmetry in spin dynamics\cite{SciRep-Tobik}.  
The energy gradient in the region between
$\mathrm{S_1}$ and $\mathrm{S_2}$ defines an effective field around which
magnetization precesses (see equations (\ref{eq-LLG}) and
(\ref{eq-Heff})). The direction of the precession is given by
gyro-magnetic moment which has a negative sign for electrons.  Thus
precession during the descent in energy between saddle points $\mathrm{S_1}$ and  $\mathrm{S_2}$ breaks the symmetry of the vortex polarization sign.  To
demonstrate this idea we have simulated several times the evolution of the
magnetization starting from various configurations between saddle points 
$\mathrm{S_1}$ and  $\mathrm{S_2}$ (see also our video\cite{supplement}).
Our calculations were performed with use of open software OOMMF\cite{OOMMF}.
The dynamics was modeled by Landau-Lifshitz-Gilbert equation with
Langevin term \cite{OOMMF-Lemcke}. We have used damping coefficient
$\alpha = 0.1$ and the amplitude of random fluctuating field was
corresponding to temperature T=300K. The trajectories in configuration
space are shown in Fig.~\ref{fig-SstateMap} (right panel) by white lines.
The precession drift towards one of the two possible vortex states 
is of course not observable in simulations, which do not contain 
precession term. In this respect simulations using just energy functional 
like e.g.~Metropolis - Hastings MC and simulations based on Landau-Lifshitz
dynamics with Langevin term differ. While MC is purely
stochastic process and therefore the choice of the minima $\mathrm{M_2}$
and $\mathrm{M_3}$ is random, the Landau-Lifshitz dynamics carries
system towards minimum $\mathrm{M_3}$ due to precession.  
With the use of our approach we thus gained additional
argument supporting dynamical origin of the symmetry breaking
mechanism in cases of particular topology of the FES.
These phenomena are likely to be observed also in other magnetic
systems. 
We believe that this precession drift during energy descent was already
observed in magnetic dots similar in shape to ours
experimentally \cite{Im-NPG} as well as numerically \cite{JMMM-Li}. 


We note that the design of CV for our system was inspired 
by the knowledge of its initial and final states \cite{PRB-Cambel, PRB-Tobik}. 
The evolution during the metadynamics exploration of the FES was smooth
indicating that the chosen set of CV is appropriate and complete. 
However, it is likely that for different systems (e.g. those allowing skyrmions)
a different choice of CV might be more appropriate. 

To summarize, we have implemented the
metadynamics algorithm to a branch of physics, where 
to our knowledge, this method was not used
before. 
This tool can enable new type of calculations in
micro-magnetism, e.g. to map the free energy landscapes
and determine the relevant barriers at finite
temperature as well as to search for new magnetic textures.
We also note that our approach may open new possibilities
of magnetization control in magnetic nano-devices.

\section*{Acknowledgement} 
We acknowledge financial support of this work by Slovak Grant Agency APVV, grant
numbers APVV-0088-12, APVV-16-0068 and to VEGA agency, grant numbers VEGA-2/0180/14,
and VEGA-2/0200/14.

\end{document}